\documentclass[traditabstract]{aa}

\usepackage{graphicx}
\usepackage{amsmath}
\usepackage{txfonts}

\usepackage{natbib}
\bibpunct{(}{)}{;}{a}{}{,}

\newcommand{\ergcm}[1]{$10^{#1}$ erg cm$^{-2}$ s$^{-1}$}

\newcommand{\hcm}[1]{$\times 10^{#1}$ cm$^{-2}$}

\newcommand{\HII}{\ion{H}{II}}

\newcommand{\ltsima}{$\buildrel < \over \sim$}
\newcommand{\lsim}{\lower.5ex\hbox{\ltsima}}
\newcommand{\gtsima}{$\buildrel > \over \sim$}
\newcommand{\gsim}{\lower.5ex\hbox{\gtsima}}

\newcommand{\xmm}{XMM-Newton}
\newcommand{\einstein}{\textit{Einstein}}
\newcommand{\chandra}{\textit{Chandra}}

\newcommand{\hess}{H.E.S.S.}
\newcommand{\msh}{MSH\,15$-$5\textit{2}}
\newcommand{\psrb}{PSR\,B1509$-$58}
\newcommand{\otevcm}[1]{$10^{#1}$ TeV$^{-1}$ cm$^{-2}$ s$^{-1}$}

\newcommand{\gr}{\ensuremath{\gamma}-ray}
\newcommand{\eemin}{E_{e,\mathrm{min}}}
\newcommand{\eemax}{E_{e,\mathrm{max}}}

\newcommand{\rs}{$R_{S}$}
\newcommand{\degree}[1]{${#1}^{\circ}$}

\begin{document}

\title{Spatially resolved \xmm\ analysis and a model of the nonthermal emission of \msh}

\author{F.M. Sch\"ock\inst{1}
\and
I. B\"usching\inst{2}
\and
O.C. de Jager\inst{2}
\and
P. Eger\inst{1}
\and
M.J. Vorster\inst{2}
}

\institute{
Erlangen Centre for Astroparticle Physics, Erwin-Rommel-Stra\ss e 1, 91058 Erlangen, Germany
\and
Unit for Space Physics, North-West University, Potchefstroom, 2520, South Africa
}

\offprints{F. M. Sch\"ock, \email{fabian.schoeck@physik.uni-erlangen.de}}

\titlerunning{Spatially resolved \xmm\ analysis and a model of the nonthermal emission of \msh}
\authorrunning{F.M. Sch\"ock et al.}

\abstract{
We present an X-ray analysis and a model of the nonthermal emission of the pulsar wind nebula (PWN) \msh. We analyzed \xmm\ data to obtain the spatially resolved spectral parameters around the pulsar \psrb. A steepening of the fitted power-law spectra and decrease in the surface brightness is observed with increasing distance from the pulsar. In the second part of this paper, we introduce a model for the nonthermal emission, based on assuming the ideal magnetohydrodynamic limit. This model is used to constrain the parameters of the termination shock and the bulk velocity of the leptons in the PWN. Our model is able to reproduce the spatial variation of the X-ray spectra. The parameter ranges that we found agree well with the parameter estimates found by other authors with different approaches. In the last part of this paper, we calculate the inverse Compton emission from our model and compare it to the emission detected with the \hess\ telescope system. Our model is able to reproduce the flux level observed with \hess, but not the spectral shape of the observed TeV $\gamma$-ray emission.
}

\maketitle

\section{Introduction}

\begin{figure}
 \centering
 \includegraphics[width=\linewidth]{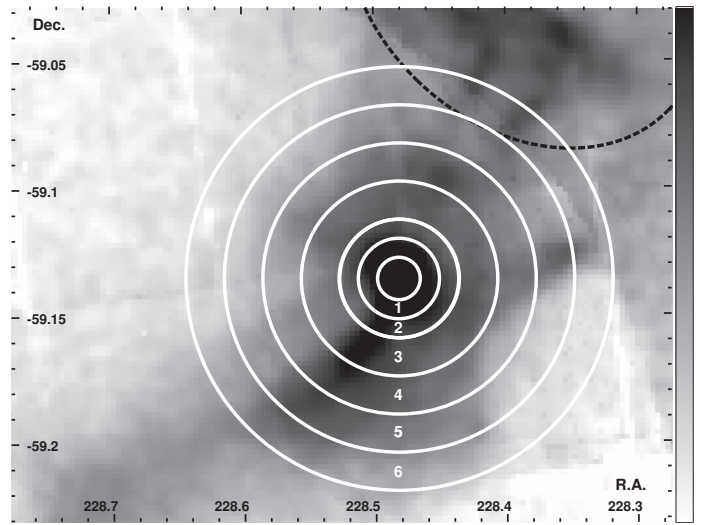}
 \caption{\xmm\ count map of \msh\ showing the six extraction regions (numbered and marked by white lines) and the exclusion region (black dashed line) used for the spectral analysis. The six extraction regions are centered on the position of the pulsar \psrb. The exclusion region was chosen to encompass the thermal emission from the \HII\ region RCW\,89.}
 \label{msh_skymap}
\end{figure}

\msh\ (G320.4$-$1.2) is a complex supernova remnant (SNR) that was discovered in radio wavelengths by \citet{1961AuJPh..14..497M}. Its radio appearance is dominated by two spots, a brighter one to the northwest and a fainter one to the southeast. The radio emission to the northwest coincides spatially with the \HII\ region RCW\,89. Situated inside the SNR is the energetic $150$\,ms pulsar \psrb, which was discovered by the \einstein\ satellite \citep{1982ApJ...256L..45S}. It has a spin-down luminosity of $\dot{E} = 1.8 \times 10^{37}\,\mathrm{erg}/\mathrm{s}$ and a characteristic age of $\tau = 1.7~\mathrm{kyr}$ \citep{1994ApJ...422L..83K}. This makes the pulsar one of the youngest and most energetic known. \citet{1999MNRAS.305..724G} concluded by a comparison of the radio and the X-ray emission that \msh, \psrb\ and RCW\,89 are associated objects and are located at a distance of $(5.2 \pm 1.4)\,\mathrm{kpc}$.

X-ray observations of the SNR have shown a pulsar wind nebula (PWN) powered by \psrb\ \citep{1996A&A...306..581T,1996PASJ...48L..33T}. Observations with the \chandra\ satellite have revealed two outflow jets in the southeast and northwest directions, the latter terminating in the optical nebula RCW\,89 \citep[henceforth referred to as G02]{2002ApJ...569..878G}. G02 derived a power-law photon index of $2.05 \pm 0.04$ and an absorption column density of $(9.5 \pm 0.3)\ $\hcm{21} for the diffuse PWN in the energy band of $0.5$-$10$\,keV. A detailed study of the innermost region of \msh\ was performed by \citet[henceforth referred to as Y09]{2009PASJ...61..129Y} using an extended data set of \chandra\ observations. Their analysis revealed hints for a ring-like feature around the pulsar, which might correspond to a wind termination shock. The hard X-ray spectrum of the PWN was observed with the BeppoSAX and the INTEGRAL satellites. The PWN is clearly seen in the off-pulse component of the emission and the morphology corresponds nicely to the measurements at lower energies. In the energy band of $20$-$200$\,keV, the off-pulse emission from the PWN is fitted best by a power law with an index of $2.1$ \citep{2001A&A...380..695M,2006ApJ...651L..45F}. 

In the very high-energy (VHE; $100\,\mathrm{GeV}$-$100\,\mathrm{TeV}$) $\gamma$-ray domain, the PWN was observed by the High Energy Stereoscopic System (\hess). The source is clearly extended beyond the point spread function (PSF) and shows a morphology comparable to what is seen in X-rays, extending northwest and southeast of the pulsar. The observed $\gamma$-ray emission is well-fitted by a power law with a photon index of $2.27$ up to a photon energy of $40\,\mathrm{TeV}$\,\citep{2005A&A...435L..17A}.

In the first part of this paper, we present an analysis of the \xmm\ data of the PWN \msh. Its large effective area makes \xmm\ ideally suited for the spectral analysis of extended sources. Our analysis thus provides a good measurement of the large-scale characteristics of \msh, compared to earlier high-resolution measurements of the inner region with the \chandra\ satellite. Following the \xmm\ analysis, we introduce a model which is based on the assumption of the ideal magnetohydrodynamic limit to describe the observed emission in the X-ray domain. Fitting the spatially resolved X-ray emission with the model, we found optimum parameters for several scenarios that have been discussed by other authors. This allows us to constrain physical quantities of the PWN termination shock and the velocity profile of the PWN. In the last part, we also apply this model to make predictions for the VHE $\gamma$-ray emission as observed by the \hess\ Cherenkov telescope array.

\section{\xmm\ Observations}

\begin{table}
\caption[]{Details of the \xmm\ EPIC PN observations on \msh.}
\label{tab-obs}
\centering
\begin{tabular}{lrr}
\hline\hline\noalign{\smallskip}
\multicolumn{1}{c}{Observation} &
\multicolumn{2}{c}{Exposures (ks)} \\
\multicolumn{1}{c}{ID} &
\multicolumn{1}{c}{performed$^{(1)}$}& 
\multicolumn{1}{c}{net$^{(2)}$} \\
\noalign{\smallskip}\hline\noalign{\smallskip}
0207050201 & 23.135 & 5.9\\
0302730201 & 16.130 & 3.6\\
0302730301 & 8.235 & 2.0\\
\noalign{\smallskip}\hline\noalign{\smallskip}
\end{tabular}
\\
$^{(1)}$ Exposure time without background screening\\
$^{(2)}$ Net exposure time after background screening
\end{table}

\begin{table}
\caption[]{Extraction regions for the \xmm\ spectral analysis of the source.}
\label{tab-reg}
\centering
\begin{tabular}{lccc}
\hline\hline\noalign{\smallskip}
\multicolumn{1}{c}{Ring} &
\multicolumn{2}{c}{Radius (arcsec)} &
\multicolumn{1}{c}{No. of Obs.}\\
\multicolumn{1}{c}{No.} &
\multicolumn{1}{c}{inner} & 
\multicolumn{1}{c}{outer}\\
\noalign{\smallskip}\hline\noalign{\smallskip}
1 & 30  & 57 & 2\\
2 & 57  & 84 & 2\\
3 & 84  & 138 & 3\\
4 & 138 & 192 & 3\\
5 & 192 & 246 & 1\\
6 & 246 & 300 & 2\\
\noalign{\smallskip}\hline\noalign{\smallskip}
\end{tabular}
\\

\end{table}

The region around the SNR \msh\ has been observed six times with \xmm\ \citep{2001A&A...365L...1J} with the EPIC-MOS \citep{2001A&A...365L..27T} and EPIC-PN \citep{2001A&A...365L..18S} cameras. These pointings were either centered north or southeast of the pulsar \psrb . During one of these pointings (Observation ID: 0312590101), all three cameras were operated in timing mode. Therefore, this observation will not be considered in the present paper. In order to study the morphology-dependent spectral characteristics of the extended emission, we require the whole area of our interest to be within the field of view (FoV). Only the three observations pointing towards the northern region matched this criterion. In each of them the detectors were operated in full-frame mode with medium optical blocking filters. The observations used in our analysis are listed in Table~\ref{tab-obs}.

For the analysis of the X-ray data we used the \xmm\ Science Analysis System (SAS) version 8.0.0 supported by tools from the FTOOLS package. For the spectral modeling, version 12.5.0 of the XSPEC software was used~\citep{1996ASPC..101...17A}. To screen the data from periods of high background-flaring activity, we used the 7 to 15\,keV lightcurve provided by the standard SAS analysis chain extracted from the full FoV. Since a good understanding of the background is crucial for the analysis of extended sources, we applied a conservative background threshold of ten background counts per second for the definition of the good time intervals (GTI). Furthermore, we only analyzed the data of the PN camera since it is more sensitive than the MOS cameras. We selected good (FLAG$=0$) single and double events (PATTERN$<=4$). The innermost region used for the spectral analysis starts at a radius of 30\,arcsec from the bright pulsar and therefore, pile-up is not an issue. All of the observations were affected by long periods of background flaring or full scientific buffer of the PN camera. This leads to rather short net exposures (see Table~\ref{tab-obs}). However, the statistics are still sufficient to obtain spectra from all of the extraction regions as defined in the next section.

\section{X-ray Spectra}
\label{section_spectra}

We extracted spectra from annular regions centered on the pulsar position (R.A.: 15:13:55, Dec.: -59:08:08.8) to study the changes of the spectral properties. The regions can be seen on the \xmm\ sky map (Fig.~\ref{msh_skymap}). We use full annuli in contrast to wedge shapes, to extract the integral flux for each radial distance. This is especially important for the modeling of the flux of the source, which is presented in Section~\ref{section_model}. The parameters for the rings are listed in Table~\ref{tab-reg}. The numbering scheme starts with ring~1, which is closest to the pulsar, and goes out to ring~6, for which the outer radius lies at a distance of 300~arcseconds from the pulsar position. Beyond the distance of 300~arcseconds the emission seems to bend sideways significantly. Since the model (see Sec.~\ref{section_model}) assumes a radial symmetry, we did not extract spectra for any regions beyond this distance.

In order to avoid systematic effects from the CCD borders, we extracted the spectra for each detector CCD separately. The background for each CCD was estimated using infield background regions from the same chip. This minimizes systematic uncertainties on the flux, compared to a background estimation from blank-sky observations. The effective areas and energy responses for each detector CCD were calculated by weighting the contribution from each pixel with the flux using a detector map from the $0.2$-$7.0$\,keV energy band. The spectra from the different CCDs for each extraction region were then merged by weighting the responses (rmf) and auxiliary responses (arf) with the size of the respective region.

\begin{figure}
 \centering
 \includegraphics[angle=270,width=\linewidth]{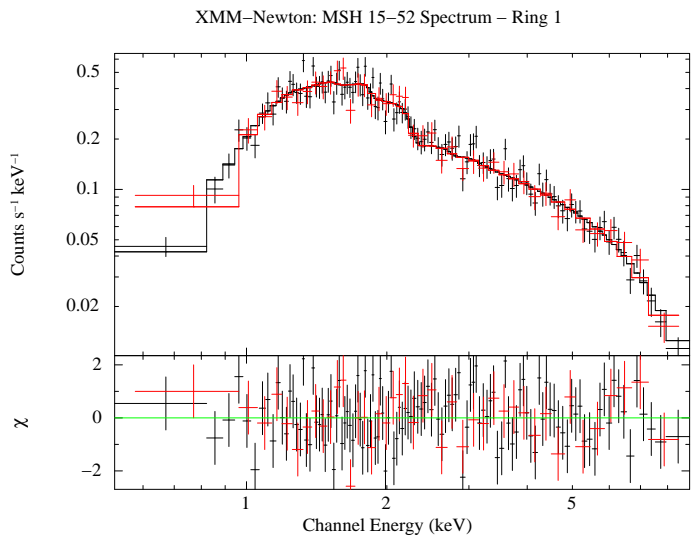}
 \caption{Spectrum and power-law fit for the first extraction region (ring~1). Two observations were used for the analysis of this region (cf. Table~\ref{tab-reg}). The resulting fit parameters are listed in Table~\ref{tab-spec}.}
 \label{msh_spec_ring1}
\end{figure}

\begin{table*}
\caption{Results obtained by fitting a power-law spectrum to the \xmm\ data of \msh.}
\label{tab-spec}
\centering
\begin{tabular}{lcccc}
\hline\hline\noalign{\smallskip}
\multicolumn{1}{c}{Ring} &
\multicolumn{1}{c}{$\Gamma$} &
\multicolumn{1}{c}{Flux} &
\multicolumn{1}{c}{Surface Brightness} &
\multicolumn{1}{c}{$\chi^{2}/\mathrm{dof}$}\\
\multicolumn{1}{c}{No.} &
\multicolumn{1}{c}{} & 
\multicolumn{1}{c}{(\ergcm{-12})} &
\multicolumn{1}{c}{(\ergcm{-17}$\mathrm{arcsec}^{-2}$)} & 
\multicolumn{1}{c}{}\\
\noalign{\smallskip}\hline\noalign{\smallskip}
1 & $1.66 \pm 0.02$ & $12.0 \pm 0.3$ & $162 \pm 4.5$ & $147/161$\\
2 & $1.78 \pm 0.03$ & $10.0 \pm 0.3$ & $92 \pm 3.0$ & $161/147$\\
3 & $1.88 \pm 0.02$ & $2.9 \pm 0.07$ & $9.2 \pm 0.2$ & $431/459$\\
4 & $1.96 \pm 0.02$ & $3.1 \pm 0.09$ & $6.1 \pm 0.2$ & $356/536$\\
5 & $2.07 \pm 0.05$ & $2.4 \pm 0.1$ & $3.4 \pm 0.2$ & $97/156$\\
6 & $2.24 \pm 0.28$ & $0.4 \pm 0.2$ & $0.5 \pm 0.3$ & $62/107$\\
\noalign{\smallskip}\hline\noalign{\smallskip}
\end{tabular}
\\

\end{table*}

For each ring, the spectra were fitted in parallel with an absorbed power-law model. In the individual observations that were used, the bad columns and the CCD borders have different orientations. This leads to a difference in the norm and thus in the flux of the fitted spectra. In the case where bad columns obscure a large fraction of the extraction region, a wrong flux is obtained. We added a selection criterion on the observations used for the spectral analysis of each ring to circumvent this problem. Only those observations were used in which the bad columns do not obscure parts of the particular extraction region. In this way we obtain the correct integral flux for each region. The resulting number of observations for each extraction region is given in Table~\ref{tab-reg}.

The parameters of the spectral power-law fit --- absorption column density, photon index and normalization --- were linked for the fitting of the different observations. The absorption column density does not vary significantly between the different regions and was fixed to the value of the first ring. The fit range was from $0.5$-$9.0$\,keV for rings~1 to 5, while for ring~6 we chose a narrower fit range of $3.0$-$9.0$\,keV, due to the insufficient statistics at low energies. As an example, Fig.~\ref{msh_spec_ring1} shows the fit of an absorbed power law to the observed emission of ring~1.

\begin{figure}
 \centering
 \includegraphics[width=\linewidth]{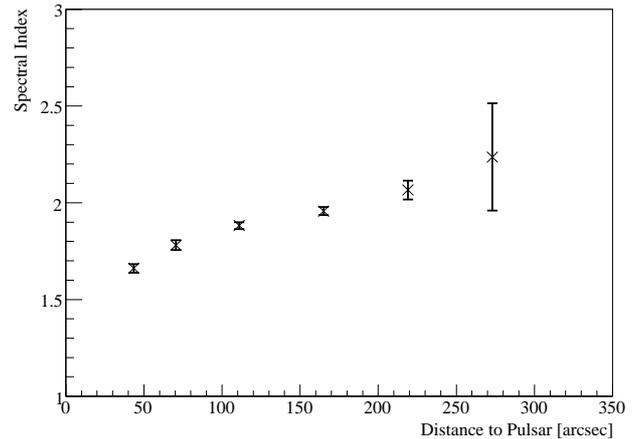}
 \caption{Spectral index of the power-law fit to the \xmm\ data of the different regions versus the mean distance of the extraction regions to the pulsar. A steepening of the spectrum is observed with increasing distance.}
 \label{fig_xmm_index}
\end{figure}

\begin{figure}
 \centering
 \includegraphics[width=\linewidth]{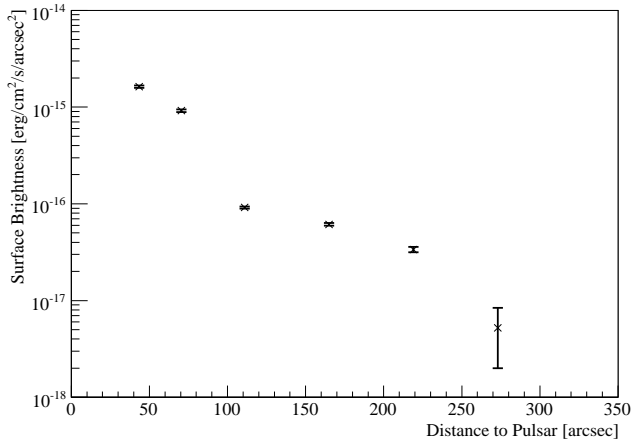}
 \caption{Surface brightness of the X-ray emission of the PWN in the $0.5$-$9.0$\,keV band as a function of the distance to the pulsar.}
 \label{fig_xmm_surfBr}
\end{figure}

The resulting parameters for the spectral analysis of each ring are listed in Table~\ref{tab-spec}. The data are fitted very well by pure power laws. Due to the increasing size of the extraction region and the constant area of the infield background, the statistical uncertainty of the spectra of the outer regions is greater, resulting in a lower value of $\chi^{2}$/dof. We obtain an absorption density of $\mathrm{N}_\mathrm{H} = (1.15 \pm 0.03)\ $\hcm{21} using the abundance tables of \citet{2000ApJ...542..914W}. This is in good agreement with results of previous X-ray analyses of this source (see e.g.~G02). The extended emission of \msh\ shows variation in the spectral properties with increasing radial distance to the pulsar. The photon index $\Gamma$ of the fitted absorbed power law increases from the inner regions to the outer regions by roughly $0.5$, reflecting a softening of the spectrum. Figure~\ref{fig_xmm_index} shows this variation of the photon index. Table~\ref{tab-spec} shows the unabsorbed flux and the surface brightness in the energy range from $0.5$-$9.0$\,keV for each extraction region as well as the $\chi^{2}$ and the number of degrees of freedom of the fit. The surface brightness decreases with increasing distance of the extraction region to the pulsar, as can be seen in Fig.~\ref{fig_xmm_surfBr}.

For the parameter optimization of the model (see Sec.~\ref{section_par_op}), we divided the spectrum for each of the rings in six intervals in energy and calculated the flux in these bins. Only four bins were used for the analysis of the ring~6 region, due to the narrower energy range. The approach of using individual flux points rather than only using the index and the total flux of a power-law fit is better suited for the parameter optimization. In this way we compare the data and the model results directly, without assuming power-law distributions for fits of the spectra.

\section{The Model}
\label{section_model}

\subsection{Injection Spectrum}

At the termination shock of the pulsar wind, particles are accelerated and injected into the PWN~\citep{2006ARA&A..44...17G}. In our model, we assume a certain particle injection spectrum at the termination shock and propagate the particles radially outward. The shape of the spectrum is chosen based on observational constraints rather than on a detailed physical model, which would be beyond the scope of our model. We assume that the particle injection spectrum $Q(E_e,t)$ at the shock radius \rs\ follows a power-law distribution. Earlier papers on the modeling of PWN concluded that the injection spectrum should follow a broken power law to account for the observed spectral properties at radio wavelengths \citep{1984ApJ...283..710K, 1984ApJ...278..630R, 2007whsn.conf...40V}:
\begin{equation}
Q(E_{e}) = 
\begin{cases}
\quad Q_{0}\left (\dfrac{E_{e}}{E_{b}} \right )^{-\Gamma_{1}} & \textrm{for} \quad E_{e} < E_{b} \textrm{,}\\
\quad Q_{0}\left (\dfrac{E_{e}}{E_{b}} \right )^{-\Gamma_{2}} & \textrm{for} \quad E_{e} \ge E_{b}\textrm{,}
\end{cases}
\label{eq_spectrum_inj}
\end{equation}
where $Q_{0}$ is a normalization constant, $E_{e}$ the lepton energy and $E_{b}$ the break energy of the spectrum. The parameters $\Gamma_{1}$ and $\Gamma_{2}$ are the spectral indices of the two power-law components. The normalization of the particle spectrum can be calculated by noting that
\begin{equation}
\int_{\eemin}^{\eemax} Q(E_e)E_edE_e = \eta \dot E\mathrm{.}
\label{eq_spectrum_norm}
\end{equation}
In this equation the parameter $\eta$ denotes the conversion efficiency of spin-down luminosity $\dot E$ of the pulsar into the energy of the leptons. The lower integration bound, corresponding to the minimum energy of the leptons, can be constrained by the observed radio spectrum and by the fact, that it has to be greater than the electron rest mass, $\eemin > 8.2 \times 10^{-7}\,\mathrm{erg}$. The maximum energy of the electrons is constrained by two conditions as specified in more detail in \citet{2008assl357deJagerDjannati}. On the one hand, the Larmor radii $R_L$ of charged particles confined within the shock radius must be less than \rs. The parameter $\epsilon$ is defined as the ratio of these two radii, $\epsilon = R_L/R_S$, and should be less than $0.5$. This limits the maximum lepton energy to
\begin{equation}
\eemax < \epsilon e \kappa \sqrt{\frac{\sigma \dot E}{c(1+\sigma)}}\mathrm{.}
\end{equation}
In this expression, $\kappa$ is the compression ratio at the shock and $\sigma$ is the magnetization parameter (ratio of the magnetic energy flux to the particle energy flux). The second condition comes into effect for strong magnetic fields. In this case the maximum lepton energy is limited by the synchrotron emission that the lepton radiates. Thus, the maximum energy is limited to
\begin{equation}
\eemax < 43.84\,B^{-1/2}\,\mathrm{erg,}
\end{equation}
where $B$ is the magnetic field strength in Gauss. The lower $\eemax$ obtained with the two limits is then used for Eq.~\ref{eq_spectrum_norm}.

Our model introduced in this article is aimed at modeling the nonthermal emission of the PWN close to the pulsar. In this region, only a young population of leptons is expected. It is therefore reasonable to neglect the time-dependence of the injection spectrum for our model. Furthermore, we focus on the modeling of the X-ray spectrum and the inverse Compton (IC) spectrum in the very high energy (VHE) $\gamma$-ray regime. For this purpose, only the power-law component beyond the break energy is relevant and we will thus assume a pure power law with $\eemin = E_b$ in the remainder of the paper. This implies that $\eta$ is only a lower limit on the conversion efficiency, since we do not take the leptons with energies below $E_b$ into account in our model.

\subsection{Propagation of the Leptons in the PWN}

In the downstream flow beyond the termination shock the particles lose energy, causing the shape of the lepton spectrum to change. In our model we use the approximation of a spherical symmetry. This implies that the propagation of the particles in the nebular flow of the PWN is strictly radial. Looking at the X-ray sky map of \msh\ (see Fig.~\ref{msh_skymap}), a deviation from a spherical symmetry is apparent. However, compared to current models in which no spatial variation of the properties of the lepton population is taken into account, the radial approach should provide a good first estimate of the effects of spatial variation in the PWN.

Assuming that the magnetic field in the PWN is toroidal, \citet{1984ApJ...283..694K} showed that the product of the flow velocity $v$, the distance from the pulsar $r$ and the magnetic field strength $B$ is a constant:
\begin{equation}
 Bvr = B_S v_S R_S = \mathrm{const.,}
\label{eq_faraday}
\end{equation}
where the subscripted parameters denote the parameters at the shock. This is valid for a steady-state solution in the ideal magnetohydrodynamic limit. The post-shocked magnetic field strength is then given by \citep{1984ApJ...283..694K,2003ApJ...593.1013S}
\begin{equation}
 B_S = \frac{\xi}{R_S} \sqrt{\frac{\dot E}{c}}\mathrm{.}
\end{equation}
In this equation, and for the parameter optimization, we use $\xi = \kappa \sqrt{\sigma / (1+\sigma)}$ as a combined parameter for the shock compression $\kappa$ and the magnetization $\sigma$.

Since we assume the approximation of a radial outflow of the leptons in the nebular flow, we can describe the bulk velocity of the leptons by a profile depending only on $v_S$ and the distance to the pulsar, $r$. We adopt a profile of the form
\begin{equation}
\label{eq_vel_profile}
 v(r) = v_S \left( \frac{R_S}{r} \right)^{\alpha}\mathrm{,}
\end{equation}
with an exponential index $\alpha$. By using this relation and Eq.~\ref{eq_faraday}, we are able to write the magnetic field as a function of the distance to the pulsar and thus it is possible to calculate the flow parameters $B$ and $v$ for every location in the PWN. With this information, we can determine the time spent by the leptons in different regions of interest throughout the nebula. Note that we neglect three aspects in our approach: the propagation of internal energy, the conservation of the total energy \citep[Eqns.~5.3 and 5.4 in][]{1984ApJ...283..694K} and the dynamical evolution of the system \citep[assuming a steady-state solution like][]{1984ApJ...283..694K}. The physics of these neglected aspects is contained in the index of the flow velocity $\alpha$ in Eq.~\ref{eq_vel_profile}. In the fit to the data we find that the spatial variation of the X-ray and TeV spectra is already adequately described by this parametrization, so that an introduction of additional free parameters would not necessarily give more information about the system.

The next step is to look at the change of the lepton spectrum as the particles propagate away from the pulsar wind shock. The particles lose energy due to adiabatic losses and synchrotron radiation in the magnetic field of the PWN. The total energy loss of the particles is given by \citep[cf.][]{1992ApJ...396..161D}:
\begin{equation}
\label{eq_losses}
\dfrac{dE}{dt} = -\dfrac{ E_{e}}{3}\mathbf{\nabla} \cdot \mathbf{v}_{\perp}(r) - 2.368 \times 10^{-3}(B_{\perp}(r)E_{e})^2\mathrm{,}
\end{equation}
where the first part represents the adiabatic losses and the second part the energy losses due to synchrotron radiation.

\subsection{Emission Spectra}
\label{section_emission}

The parameters of the lepton population and the magnetic field are described for all locations in the PWN by the Eqs.~\ref{eq_faraday} to \ref{eq_losses}. With this information we are able to calculate the spectra of the synchrotron and inverse Compton radiation emitted by the leptons. For the calculation of the two emission processes, we use the standard equations by \citet{1970RvMP...42..237B}.

We are now able to calculate the total emission for the lepton population at each point in the PWN. The emission from the model calculations is summed up to match the regions as observed in the \xmm\ analysis (see Sec.~\ref{section_spectra}). By comparing the measured synchrotron spectra with the calculated spectra we are able to constrain the free parameters of our model.

\section{Parameter Optimization}
\label{section_par_op}

The optimization of the parameters of the model, as introduced in Section~\ref{section_model}, was carried out by dividing the PWN into a number of shells. As discussed in the previous section, the values of $v$ and $B$ are known for every location in the PWN, thus enabling us to calculate the amount of energy lost by a particle during its propagation from the shock up to a specified shell. This, in turn, allows us to calculate the lepton spectrum that enters a shell, as well as the change in the lepton spectrum as it propagates through the shell. To calculate the corresponding nonthermal spectra, the initial lepton spectrum is used. This is sufficient, provided that the size of the shells is chosen small enough. The modified spectrum is then used as the injection spectrum for the following shell. For the first shell, the lepton spectrum is calculated by making use of Eq.~\ref{eq_spectrum_inj}, i.e. the lepton injection spectrum at the shock.

Since our modeling focuses only on the X-ray synchrotron component, we use a single power law, with $E > E_b$, for the lepton injection spectrum. The index of the lepton spectrum is derived from measurements of the photon spectrum close to the termination shock with the \chandra\ X-ray telescope, which yielded a photon index of around $1.5$ (Y09). Assuming that the lepton population emitting this spectrum has not undergone any significant cooling, the spectral index of the lepton spectrum is then equal to $2$, which we adopt for our optimizations. G02 concluded that the spectral break between the comparatively flat radio spectrum and the steeper synchrotron spectrum is just below the X-ray band. Thus it is reasonable to assume $\eemin$ to be of order $1$\,erg. Since only the logarithm of $\eemin$ contributes to the normalization of the spectrum, a variation of $\eemin$ does not change the spectral shape and has little effect on the total flux.

The remaining free parameters of the model are then the shock radius \rs, the bulk velocity of the leptons at the shock $v_{S}$, the index of the velocity profile $\alpha$, the conversion efficiency of spin-down luminosity into lepton energy $\eta$ and the parameter $\xi$, which links the magnetization $\sigma$ and the compression $\kappa$ of the shock. The parameters \rs\ and $v_S$ are restricted by the observations of G02 and Y09, whereas the other parameters are left unconstrained for the optimization. To determine the optimum fit to the synchrotron spectra for the different scenarios, we minimized the $\chi^2$ test statistic for the \xmm\ data points and the calculated model flux points.

For \psrb, G02 and Y09 both estimated the termination shock radius using \chandra\ data, but do not find consistent results. G02 favor a scenario in which a feature at $0.5$\,pc corresponds to an internal structure in the termination shock. They also estimated the $\sigma$ value for several compact knots of emission less than $0.5$\,pc away from the pulsar and concluded that the transition to a particle-dominated wind should occur at a distance less than $0.1$\,pc from the pulsar. Y09 analyzed a considerably larger data set of \chandra\ observations and used image enhancement techniques to resolve the detailed structure of the inner region around \psrb. According to their results, the termination shock is located at a distance of $9\,''$ to the pulsar, which corresponds to a termination shock radius of $R_S = 0.225$\,pc. For the magnetization parameter, Y09 derived a value of $\sigma \approx 0.01$, which is about a factor of $2$ greater than the result obtained by G02.

Based on the two analyses, we considered three scenarios with different values for \rs\ for the optimization procedure of our model. In Scenario~I we assume that the termination shock is unresolved in the \chandra\ observations by G02 and Y09 and thus is at a distance closer than $0.1$\,pc from the pulsar. For Scenario~II we assume a distance of $R_S = 0.5$\,pc, while for Scenario~III we adopt the value of $R_S = 0.225$\,pc, as stated by Y09.

G02 and Y09 both assume a shock velocity of $v_S \approx c/3$. The same value was also found for other PWN, e.g. the Crab Nebula \citep{1984ApJ...283..694K}. Furthermore, this shock velocity is in good agreement with the lower limit that G02 determined from the energetics of the PWN. Therefore, we also adopted this value for our calculations.
We do not constrain the remaining parameters $\alpha$, $\eta$ and $\xi$, except for the straight-forward assumption that the conversion efficiency $\eta$ should be in the interval $0 < \eta < 1$. Table~\ref{tab-modelpars} gives an overview of the three scenarios and the constraints on the parameters, which we used for the modeling described in the next section.

\begin{table}
\caption[]{Model parameters for the different scenarios as discussed in the text.}
\label{tab-modelpars}
\centering
\begin{tabular}{ | l | c | c | c | }
\hline
\multicolumn{1}{ | l | }{Parameter} &
\multicolumn{3}{ | c | }{Scenario}\\ \cline{2-4}
\multicolumn{1}{ | c | }{} &
\multicolumn{1}{ c | }{I} & 
\multicolumn{1}{ c | }{II} & 
\multicolumn{1}{ c | }{III}\\
\hline
\rs\ [pc] &  $< 0.1$ & $0.5$ & $0.225$\\
\hline
$v_S$ [$c$] &
\multicolumn{3}{ c |}{$1/3$}\\
\hline
$\alpha$ &
\multicolumn{3}{ c |}{no constraint}\\
\hline
$\eta$ &
\multicolumn{3}{ c |}{$0 - 1$}\\
\hline
$\xi$ &
\multicolumn{3}{ c |}{no constraint}\\
\hline
\end{tabular}
\\

\end{table}

\section{Results of the Modeling}
\label{section_results}

Using the model and the parameter constraints described in the previous section, we calculated the optimum parameters for the three scenarios defined in Section~\ref{section_par_op}. The results of the optimization show that the Scenarios~II and III (larger shock radius) yield a better fit to the data than Scenario~I (small shock radius). For Scenario~I we do not have a fixed value for \rs. We thus leave it as a free parameter for the simulations. The best fit for this scenario is obtained for a value of $R_S = 0.1$\,pc which is the upper limit for \rs\ (see Table~\ref{tab-modelpars}). However, even the best fit of Scenario~I does not give a satisfactory result. Due to the small shock radius, the synchrotron cooling is very efficient and leads to a strong cutoff in the X-ray spectra for the outer rings, which is not consistent with the observed emission.

\begin{figure}
 \centering
 \includegraphics[width=\linewidth]{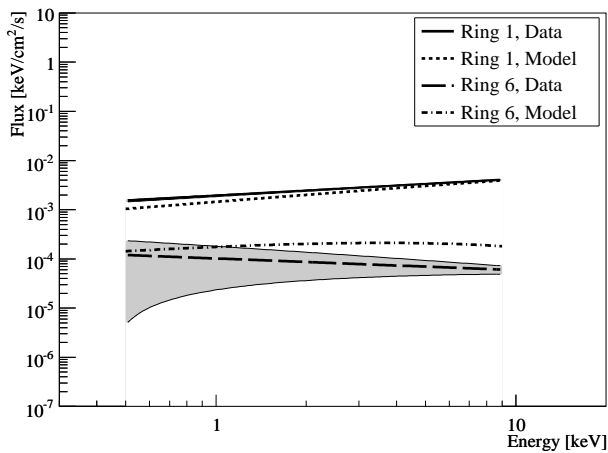}
 \caption{Data and model spectra for the ring~1 and 6 regions for Scenario~II. The shaded regions mark the error bands of the \xmm\ measurement of ring~1 and 6. For ring~1 the error band is very narrow and thus hardly visible.}
 \label{fig_sc2_spectra}
\end{figure}

\begin{figure}
 \centering
 \includegraphics[width=\linewidth]{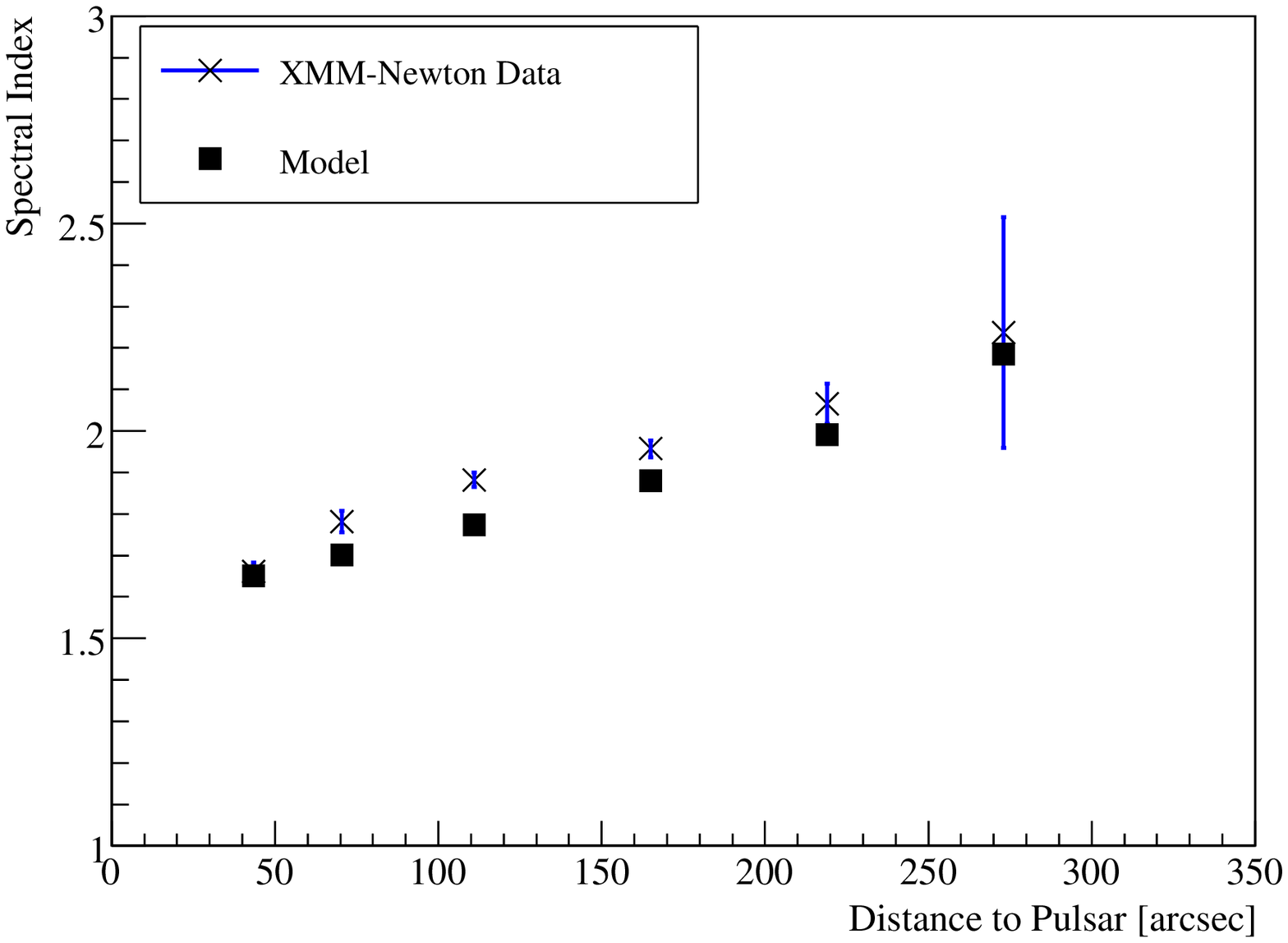}
 \caption{Spectral indices of power-law fits to the \xmm\ data of the different regions in the energy range of $0.5$-$9.0$\,keV. The model points shown are calculated for the best fit of Scenario~II and use the same energy range for the fit.}
 \label{fig_sc2_index}
\end{figure}

\begin{figure}
 \centering
 \includegraphics[width=\linewidth]{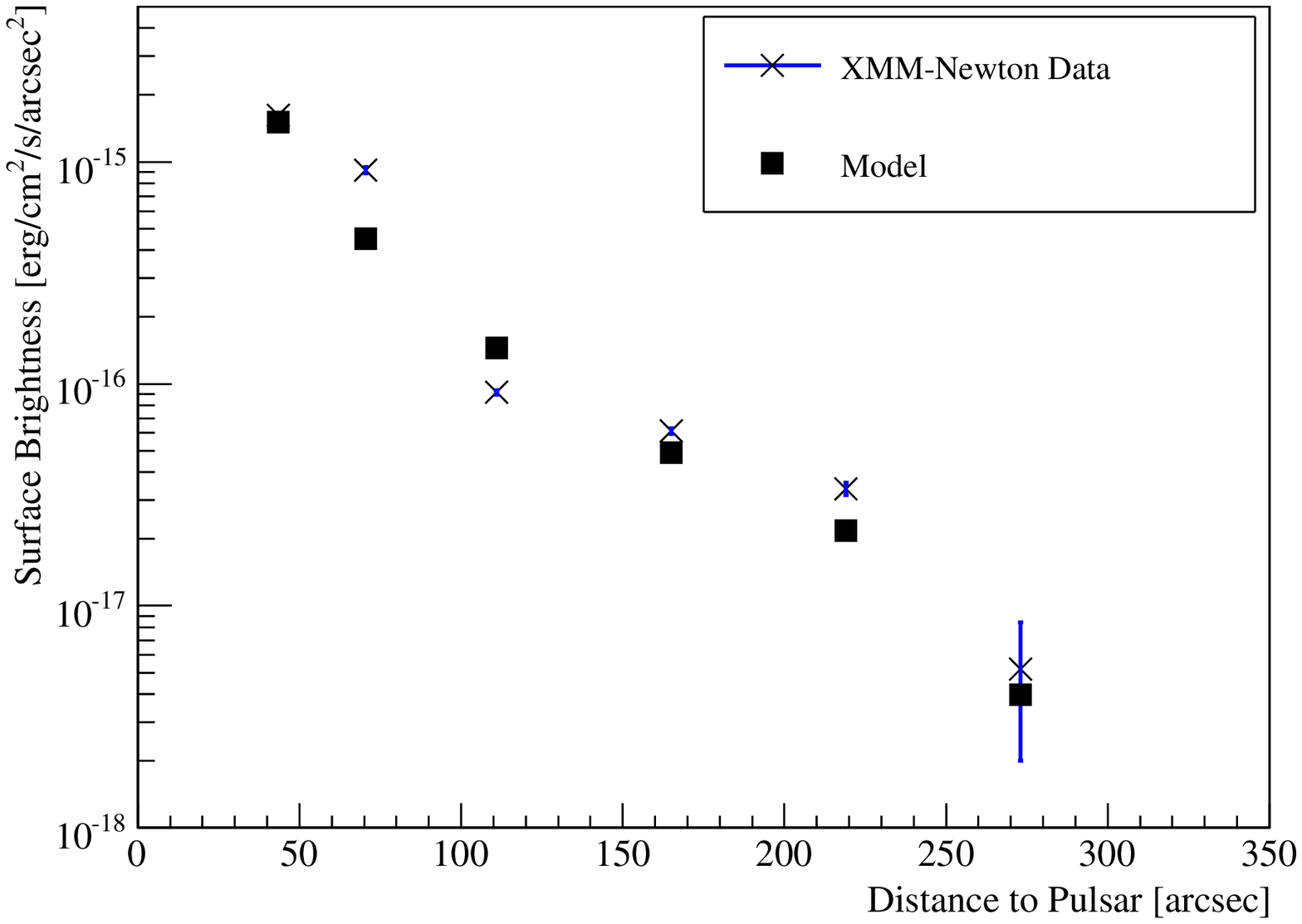}
 \caption{Variation of the surface brightness for the \xmm\ data and the model in the energy range of $0.5$-$9.0$\,keV. The model points shown are calculated for the best fit of Scenario~II.}
 \label{fig_sc2_surfBr}
\end{figure}

\begin{figure}
 \centering
 \includegraphics[width=\linewidth]{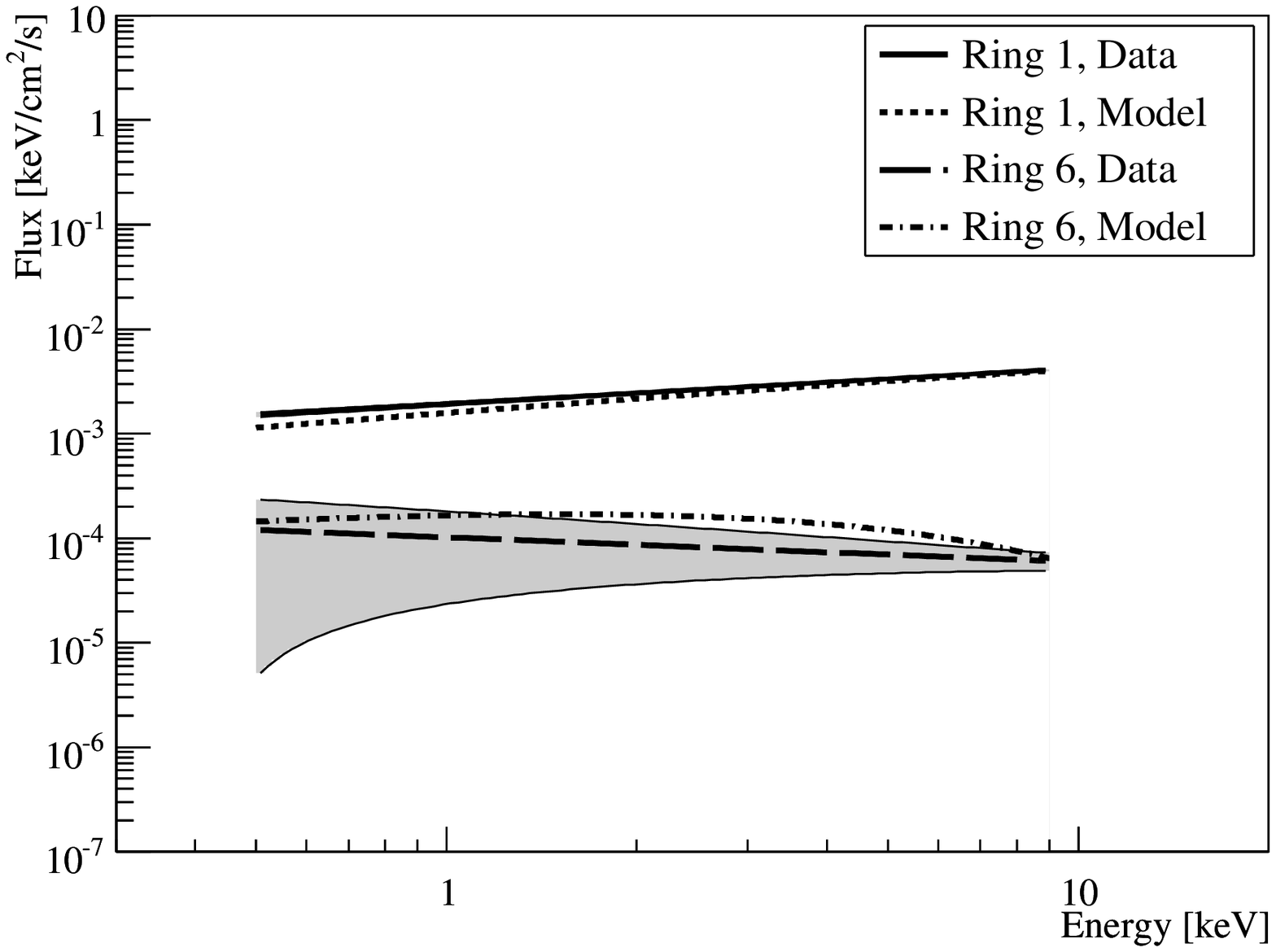}
 \caption{Data and model spectra for the ring~1 and 6 regions for Scenario~III. The shaded regions mark the error bands of the \xmm\ measurement of ring~1 and 6. For ring~1 the error band is very narrow and thus hardly visible.}
 \label{fig_sc3_spectra}
\end{figure}

\begin{figure}
 \centering
 \includegraphics[width=\linewidth]{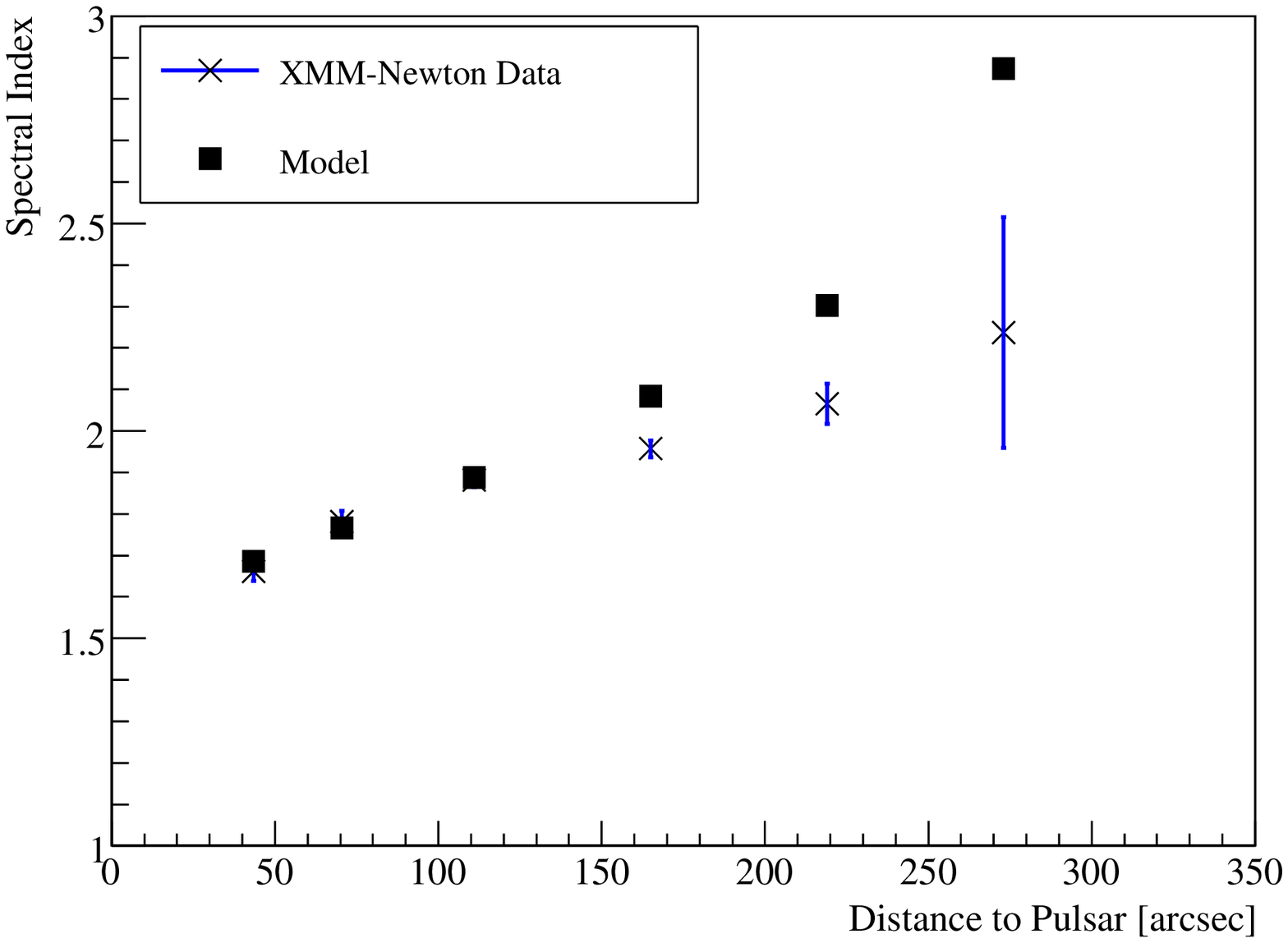}
 \caption{Spectral indices of power-law fits to the \xmm\ data of the different regions in the energy range of $0.5$-$9.0$\,keV. The model points shown are calculated for the best fit of Scenario~III and use the same energy range for the fit.}
 \label{fig_sc3_index}
\end{figure}

\begin{figure}
 \centering
 \includegraphics[width=\linewidth]{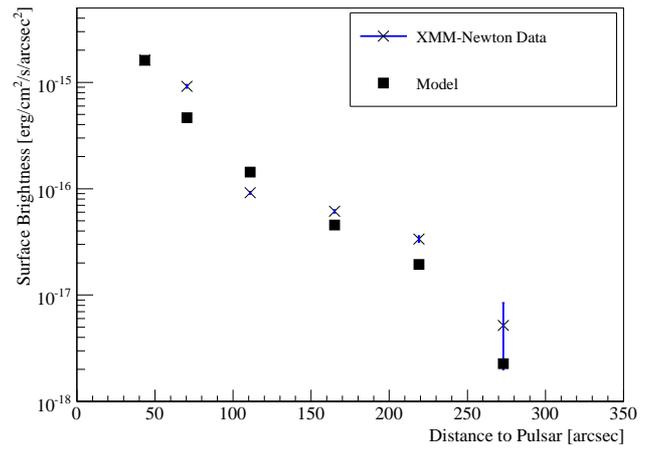}
 \caption{Variation of the surface brightness for the \xmm\ data and the model in the energy range of $0.5$-$9.0$\,keV. The model points shown are calculated for the best fit of Scenario~III.}
 \label{fig_sc3_surfBr}
\end{figure}

The results obtained for Scenario~II yield a better fit to the data. In this scenario the shock radius is considerably larger. Thus, the leptons accelerated at the termination shock suffer from less synchrotron cooling up to ring~1 of the \xmm\ measurement. Figure~\ref{fig_sc2_spectra} shows as an example the data and model synchrotron spectra for rings~1 and 6 for the optimum set of parameters found in Scenario~II. Even the best-fit model spectrum deviates significantly from the \xmm\ spectrum, but this is expected due to the simplifications of our model. The general trend, however, reproduces the spectral shape and the observed flux level. The spectral index with increasing distance of the extraction region from the pulsar is plotted in Fig.~\ref{fig_sc2_index}. The best-fit model is able to reproduce the variation of the spectral index. As shown in Fig.~\ref{fig_sc2_surfBr}, the change in surface brightness with increasing distance of the extraction regions to the pulsar is also reproduced. For Scenario~III we also get a good fit to the data. Figure~\ref{fig_sc3_spectra} shows the spectra for the optimum set of parameters found in Scenario~III, again for rings~1 and 6. The spectral index and the surface brightness with increasing distance can be seen in Figs.~\ref{fig_sc3_index} and \ref{fig_sc3_surfBr}. The shape is also reproduced well for this scenario.

\begin{figure}
 \centering
 \includegraphics[width=\linewidth]{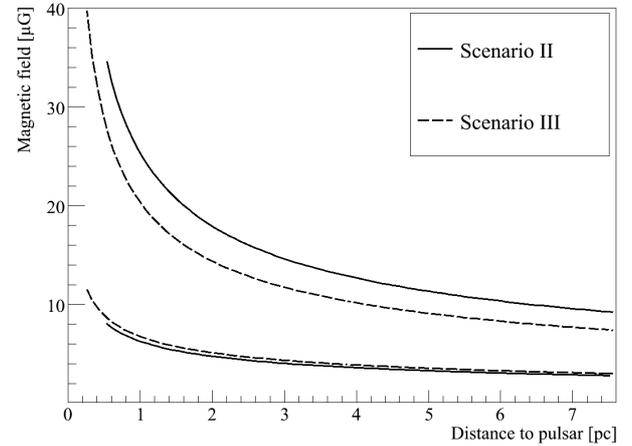}
 \caption{Spatial evolution of the magnetic field with distance from the pulsar. The two solid lines denote the upper and lower range of values of the magnetic field strength for Scenario~II (based on the range for $\alpha$ and $\xi$ that is found in the optimization). The two dashed lines mark the upper and lower limit of $B$ for Scenario~III.}
 \label{fig_Bfield}
\end{figure}

The results obtained for Scenarios~II and III constrain the range of the model parameters. Since the parameters are correlated, it is only possible to state parameter ranges. For Scenario~II we found that the index of the velocity profile is in the range of $\alpha = 0.4 - 0.6$. The conversion efficiency is greater than $0.3$ for this scenario and $\xi$ ranges between $0.3$ and $1.3$. The compression ratio $\kappa$ varies between $1$ and $3$ for relativistic shocks. We can thus translate the range of $\xi$ to the result that $\sigma > 0.01$, which is a factor of $2$ more than the lower limit derived by G02. For Scenario~III, we found the same range for $\alpha$ and $\eta$, but a narrower lower limit on the magnetization parameter, $\sigma > 0.005$. This is lower than the limit that Y09 found in their estimate based on the equipartition assumption. In summary, we can state that our results support a shock radius of the order of $0.2 - 0.5$\,pc. However, we are not able to favor one scenario above the other. For the conversion efficiency, we found a lower limit of $\eta > 0.3$ for Scenario~II and III. This agrees very well with the result of the time-dependent one-zone model by \citet{2008ApJ...676.1210Z}. The magnetic field estimates for \msh\ range between $8$\,$\mu$G (G02) and $25$\,$\mu$G~\citep{2008ApJ...676.1210Z}. The spatial evolution of the magnetic field with distance for our model can be seen in Fig.~\ref{fig_Bfield}. Our predictions for the magnetic field in the PWN agree with the estimates from the other authors.
The lower limit of $\sigma > 0.005$ on the magnetization parameter agrees very well with the observational results obtained with the \chandra\ satellite by G02 and Y09, which both state a lower limit of the same value as we derive with our model.

\section{Implications for the TeV $\gamma$-ray Emission}
\label{section_tev}

\begin{figure}
 \centering
 \includegraphics[width=\linewidth]{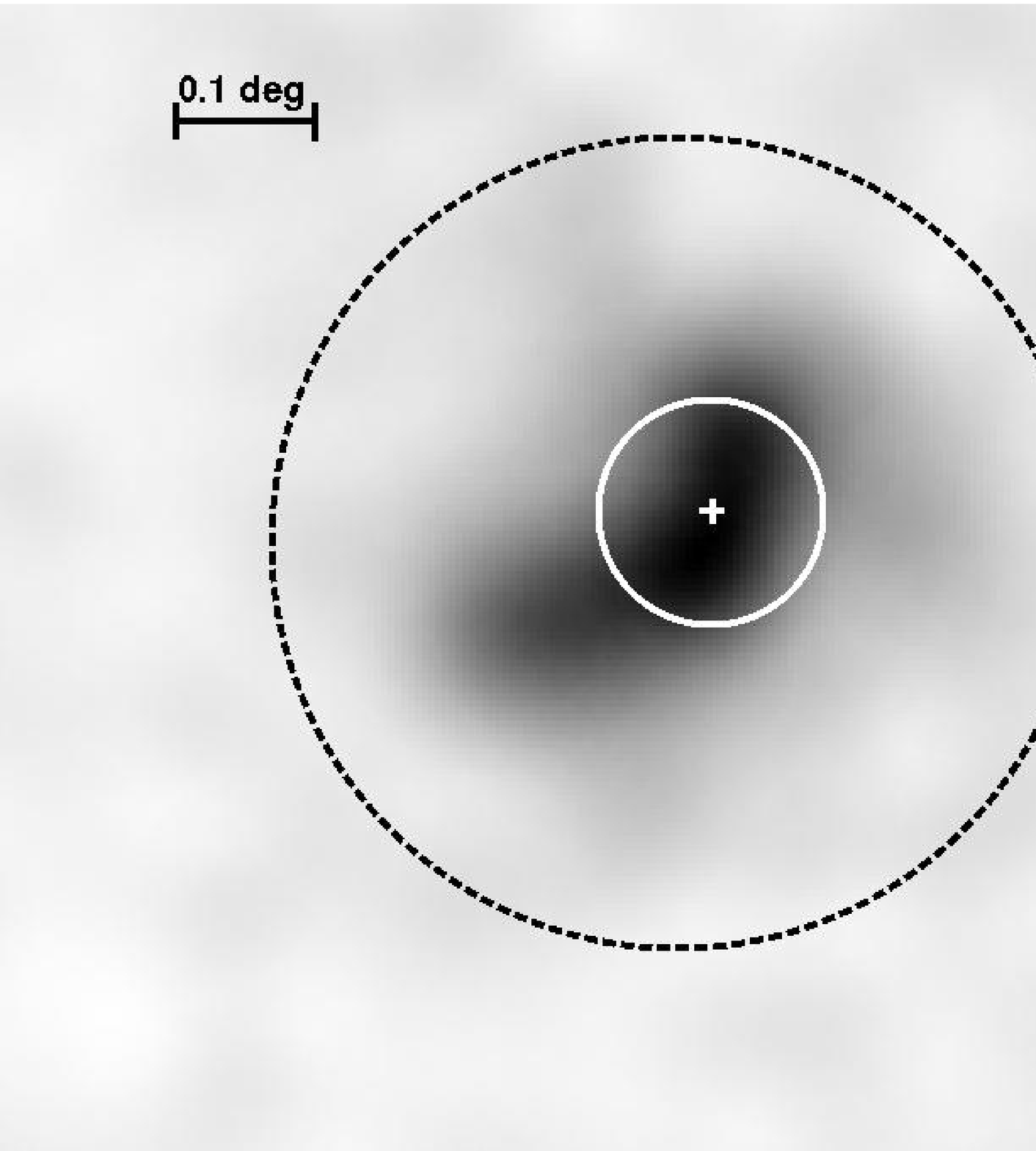}
 \caption{Smoothed \hess\ excess map of the VHE $\gamma$-rays from \msh\ \citep{2005A&A...435L..17A}. The map was smoothed by a Gaussian with the size of the point spread function ($\approx$\degree{0.07}) to reduce statistical fluctuations. The extraction region for the published \hess\ spectrum is marked with a dashed black line, the white solid line denotes the $300$\,arcsec region used for the modeling in this paper. The position of the pulsar is indicated by a white cross.}
 \label{fig_hess_skymap}
\end{figure}

\begin{figure}
 \centering
 \includegraphics[width=\linewidth]{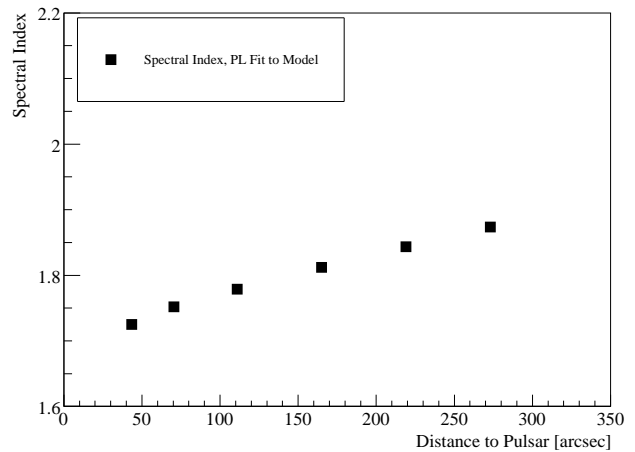}
 \caption{Spectral index of a power-law fit to the IC emission calculated in the model with increasing distance from the pulsar. The fit range was from $300$\,GeV to $30$\,TeV. Plotted is the result of the model with the best fit parameters for Scenario~II for a conversion efficiency of $\eta = 0.4$.}
 \label{fig_IC_index}
\end{figure}

Based on the optimized parameters described in the previous section, we calculated the IC emission for our model to compare it to the observational TeV $\gamma$-ray data. \msh\ is an extended source in VHE $\gamma$-rays and was detected by the \hess\ experiment \citep{2005A&A...435L..17A}. The emission in the VHE energy band extends beyond the central part of the PWN that is considered in our model. Thus, we used the published VHE excess events map to rescale the spectrum to the central $300$\,arcsec region around the pulsar. The excess map, smoothed with a Gaussian with the size of the point spread function ($\approx $\degree{0.07}, is shown in Fig.~\ref{fig_hess_skymap}. The ratio of excess events between the two regions used in the \hess\ publication and for our modeling is $3.42$. The rescaled normalization of the \gr\ spectrum at $1$\,TeV is then $(1.67  \pm 0.06_{\textrm{stat}}  \pm 0.41_{\textrm{sys}}) \times\ $\otevcm{-12}, where we rescale the errors with the same factor. Our rescaling assumes that there is no significant spectral variation for the \hess\ spectrum, i.e. the photon index of the measured index is equal to $2.27 \pm 0.03_{\textrm{stat}} \pm 0.2_{\textrm{sys}}$ throughout the extraction region. Another important aspect to consider is the size of the \hess\ PSF. Since the extension of its $68\%$ containment radius is of the same order as our extraction region ($R_{68} \lsim 0.1$\,deg), we expect some contamination in our extraction region from the outer regions of the PWN. About $30\%$ of the emission from inside the PSF will fall outside of the region defined by the rings~1 to 6. However, a similar amount from the emission outside the PSF region will fall into the central PSF region (i.e. ring~1 to 6). These effects will cancel to a first order, although a systematic steepening in the spectrum is expected if the TeV photon index also steepens with increasing radius, as is predicted by our model. Figure~\ref{fig_IC_index} shows the spectral index for the IC emission with increasing distance of the region from the pulsar, where it can be seen that the index steepens by $\Delta \Gamma \approx 0.15$.

\begin{figure}
 \centering
 \includegraphics[width=\linewidth]{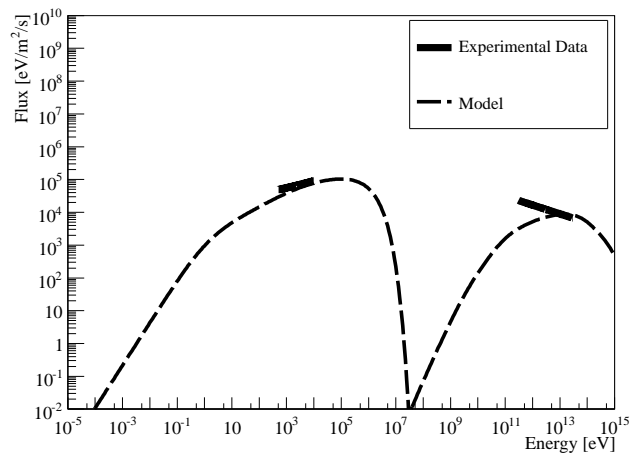}
 \caption{Spectral energy distribution of \msh\ from the extraction regions as marked in Fig.~\ref{fig_hess_skymap}. The \xmm\ and the \hess\ data are marked as solid lines, the model results for the best fit values of Scenario~II with a conversion efficiency of $\eta = 0.4$ are drawn as dashed lines.}
 \label{fig_SED}
\end{figure}

Figure~\ref{fig_SED} shows a plot of the spectral energy distribution (SED) of \msh\ in the X-ray and TeV \gr\ band. The presented experimental data in the X-ray band are the sum of the emission of the shells from the \xmm\ analysis (see Sec.~\ref{section_spectra}). For the TeV range, the rescaled \hess\ spectrum is shown. The model SED was obtained with the best fit parameters of Scenario~II (\rs$ = 0.5$\,pc) for a conversion efficiency of $\eta = 0.4$. For this set of parameters, the fitted spectral index to the IC emission in the $0.3$-$30$\,TeV energy range is $1.8$ and thus slightly below the statistical and systematic errors of the measured \hess\ spectrum. The normalization of the fit is $3.1 \times\ $\otevcm{-12} and hence also below the $1 \sigma$ errors of the measurement. The ratio of the predicted flux above $0.3$\,TeV for this set of parameters to the observed flux is $0.3$.

\begin{figure}
 \centering
 \includegraphics[width=\linewidth]{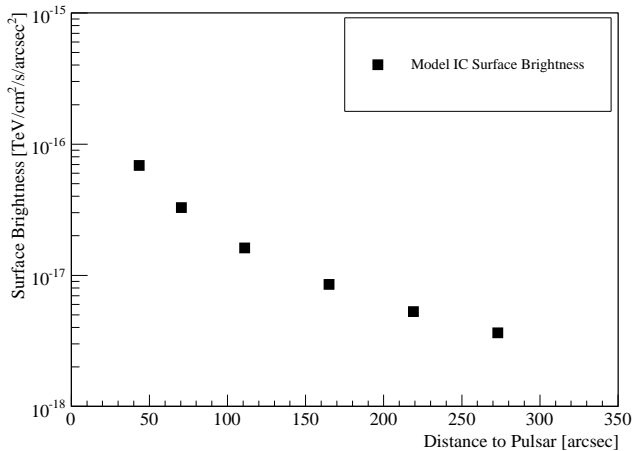}
 \caption{Surface brightness of the IC emission with increasing distance from the pulsar. The surface brightness was calculated for a power-law fit to the data in an energy range of $300$\,GeV to $30$\,TeV. Plotted is the emission from the model calculations with the best fit parameters for Scenario~II for a conversion efficiency of $\eta = 0.4$.}
 \label{fig_IC_surfBr}
\end{figure}

The change in surface brightness of the IC emission is shown in Fig.~\ref{fig_IC_surfBr}. The flux was calculated in the energy interval of $0.3$-$30$\,TeV and for the same ring regions as used in the X-ray analysis. Unfortunately, with the angular resolution of the \hess\ experiment it is not possible to test the model against the data. Future Imaging Air Cherenkov Telescopes like the Cherenkov Telescope Array \citep[CTA,][]{2009arXiv0912.3742W} will have a resolution around one arcminute and will allow for a test of these predictions. Compared to the X-ray surface brightness, which drops by three orders of magnitude from ring~1 to ring~6, the drop in the surface brightness of the IC emission is considerably smaller. However, from looking at the VHE sky map (Fig.~\ref{fig_hess_skymap}) one can see, that the emission must have a large component which is not described by our model. This might, for example, be due to an older population of electrons.

\section{Conclusion}
\label{section_conclusion}

We analyzed \xmm\ data of the SNR \msh\ to derive the spatially resolved spectral parameters of the inner region of the source. For this analysis we extracted spectra from six annuli centered on \psrb. A steepening of the X-ray spectrum with increasing distance from the pulsar is observed. The surface brightness in the \xmm\ range drops by three orders of magnitude from the inner region close to \psrb\ up to the last ring used in our analysis (at a distance of $300$\,arcsec from the pulsar). The spectra of all rings are fitted well with power laws.

We then introduced a numerical model to describe the spatial evolution of the lepton population and the magnetic field within the PWN. Our model is based on the parameters of the termination shock. We fitted the calculated model spectra to the \xmm\ data to constrain the parameters of our model. The results of our optimizations suggest a termination shock in the range of $0.225$ to $0.5$\,pc (Scenarios~II and III in this work), which are the values derived from different analyses by G02 and Y09. However, the results of the optimizations of our model do not favor one scenario over the other. For the magnetization parameter, we arrive at a lower limit of $0.005$, while our lower limit on the conversion efficiency is $0.3$.

Using the model parameters obtained from the fit to the X-ray data, we calculated the IC emission from the lepton population in the PWN. We rescaled the published \hess\ data  to compare the total IC emission from our model with the observed emission. The results show that our model reproduces the right order of magnitude for the observed IC flux, but differs slightly in the spectral shape of the emission. However, we did not take into account contamination effects for the VHE photons due to the PSF. A steeper photon index outside the region used for our model would lead to a systematically steeper spectrum inside the region due to PSF spillover effects. This effect was not taken into account in Fig.~\ref{fig_SED}. Additionally, we also calculated the spectral index and the surface brightness with increasing distance of the regions from the pulsar, which may be tested with future VHE $\gamma$-ray instruments like the CTA observatory.

The total energy that is emitted by the leptons in the inner region of the PWN via synchrotron and IC emission in our model can be obtained by integrating over the SED (see Fig.~\ref{fig_SED}). The integrated emission of $7.6 \times 10^{35}\,\mathrm{erg\,s}^{-1}$ amounts to roughly $4\%$ of the current spin-down luminosity of \psrb. However, when making this comparison one has to be aware that the integrated emission comes from a population of leptons with different lifetimes and cannot be directly linked to the current spin-down luminosity. Furthermore, there is also significant emission at X-ray and TeV energies outside the central \degree{0.1} region considered in our modeling approach, which increases the percentage of lepton energy converted to radiation (see Figs.~\ref{msh_skymap} and \ref{fig_hess_skymap}).

In summary, we show that our radially symmetric model of the PWN \msh\ readily describes the nonthermal emission seen in the X-ray band. Despite the simplifying assumptions of our model, the change in spectral index and surface brightness can be reproduced. The model is also able to predict the flux level of the observed emission in VHE $\gamma$-rays, however, the exact shape of the spectrum is not reproduced. For the future, an application of the model introduced in this paper to other PWN should be interesting to obtain a broader view on the underlying physics in PWN.

\acknowledgements{
We would like to thank the anonymous referee for the constructive comments. They helped to improve the article significantly.
}

\bibliographystyle{aa}
\bibliography{14151}

\end{document}